# Time-resolved x-ray microscopy for materials science

Haidan Wen[1], Mathew J. Cherukara[2], and Martin V. Holt[2]

[1] Advanced Photon Source, Argonne National Laboratory, Argonne, IL 60439;
  email: wen@anl.gov
[2] Center for Nanoscale Materials, Argonne National Laboratory, Argonne, IL 60439;
  email: mcherukara@aps.anl.gov; mvholt@anl.gov

**Abstract**

X-ray microscopy has been an indispensable tool to image nanoscale properties for materials research. One of its recent advances is to extend microscopic studies to the time domain for visualizing the dynamics of nanoscale phenomena. Large-scale x-ray facilities have been the powerhouse of time-resolved x-ray microscopy. Their upgrades including a significant reduction of the x-ray emittance at storage rings and fully coherent ultrashort x-ray pulses at free electron lasers, will lead to new developments in instrumentation and open new scientific opportunities for x-ray imaging of nanoscale dynamics with the simultaneous attainment of unprecedentedly high spatial and temporal resolutions. This review presents recent progress in and the outlook for time-resolved x-ray microscopy in the context of ultrafast nanoscale imaging and its applications to condensed matter physics and materials science.

# 1: Introduction

As systems scale up from a single atom to millions of atoms and, eventually, to macroscopic real-world materials, the interactions among individual objects increase exponentially and standard theoretical approaches lose the ability to accurately predict the materials behaviors in this regime. In many such systems, the individual particles become less important, while the collective behavior emerges to play a critical role. The competition among multiple collective states leads to coexisting states (heterogeneities) with distinct structural, electronic or spin configurations at cross-over length scales from nm to µm, so that the governing physics transits from quantum to classical physics (1).

"More is different," as P. W. Anderson summarized (2). For example in solid-state materials, the intrinsic dynamics of heterogeneity is intimately tied to a number of fundamental phenomena, such as metal-to-insulator phase transitions (3), high-temperature superconductivity (4), and colossal magnetoresistance (5). These emergent properties not only occur under thermal equilibrium but can also be created dynamically when the system is driven into a nonequilibrium state (6, 7). However, en route to a deeper understanding of these fundamental and emergent phenomena, heterogeneity is often a major obstacle to connecting real-world inhomogeneous dynamics with theories that are developed for a simplified model consisting homogenous collection of single particles. Visualizing the evolution of heterogeneities in this cross-over regime of time and space with compatible spatiotemporal resolution is important for understanding, developing, and controlling novel material properties. Experimental characterization of the complex landscape of heterogeneous structures from nano to mesoscales and their evolution in time represents one of the grand challenges in modern materials science (1).

To capture the evolution of these heterogeneous structures, spatiotemporally resolved characterization tools are needed. Historically, they are developed independently to address the resolution requirement in either spatial or temporal domains. Tremendous efforts have been exerted to achieve ultrahigh spatial resolution down to femtometer scales, and temporal resolution down to attosecond scales, providing unprecedented precision in visualizing processes at the levels of electrons and atoms.

In the spatial domain, there are a variety of widely used imaging tools, for imaging in length scales ranging from atomic to mesoscale resolution (Figure 1a). Each of them is sensitive to one or several degrees of freedom including lattice, spin or electronic orders. Real-space imaging modalities include (a) scanning probes using sharp tips such as scanning tunneling microscopy, atomic force microscopy, and piezoresponse force microscopy or focused radiation such as optical microscopy, x-ray microscopy, and electron microscopy; (b) non-scanning probes that image a probing area onto a detector such as full-field imaging, or utilizes the coherence of the radiation to reconstruct the real-space density distribution such as coherent imaging. Imaging tools also can operate in the reciprocal spaces. Through Fourier transform, a comprehensive characterization of reciprocal space is equivalent to a measurement of materials in real space. In some cases, both real space and reciprocal space imaging are employed so that the real space heterogeneities on nanoscales is linked to crystallographic ordering with atomic resolution, such as bright-field imaging in electronic diffraction (8), scanning x-ray diffraction microscopy (9, 10), and x-ray ptychography (11).

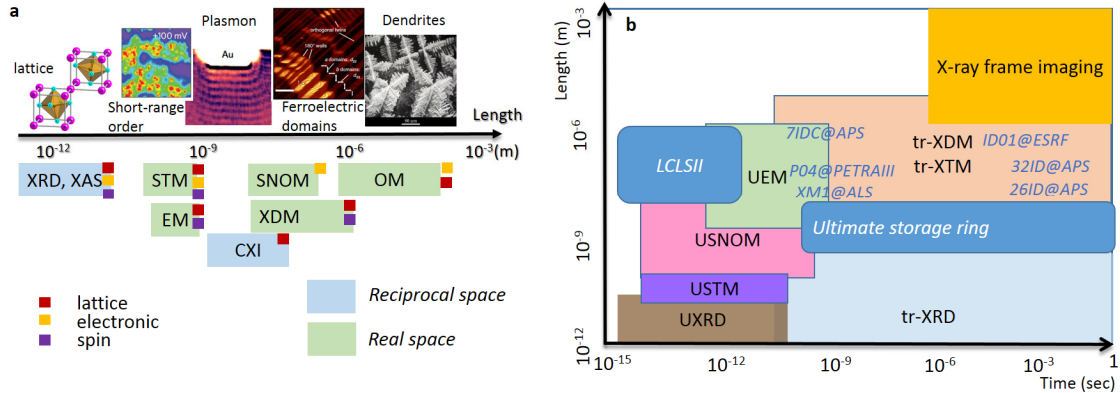

Figure 1 (a) The length scale with the corresponding phenomena and characterization techniques. XRD: x-ray diffraction, XAS: x-ray absorption spectroscopy, STM: scanning tunneling microscopy, EM: electron microscopy, SNOM: scanning near-field optical microscopy; OM: optical microscopy; XDM: x-ray diffraction microscopy; CXI: coherent x-ray imaging. The color squares represent the degrees of freedom that the imaging techniques are sensitive to measure. The color rectangles represent the measurement in real or reciprocal spaces. (b) The length and time scales that are accessible by various spatiotemporally resolved imaging techniques. New facilities are marked with rounded rectangles and representative beamlines are labeled by the blue letters. "U" represents "ultrafast" and "tr" represents "time-resolved". XTM: x-ray transmission microscopy. APS: Advanced Photon Source, ALS: Advanced Light Source, LCLS-II: Linac Coherent Light Source II, ESRF: European Synchrotron Radiation Facility.

In the temporal domain, the pump-probe technique can measure deterministic processes with a time resolution determined by the pulse duration, rather than the detector speed in conventional multiframe imaging technique. The low-cost high-power ultrafast laser systems can deliver ultrafast pulses to enable dynamical studies on unprecedented femtosecond time scales (12). Many intrinsic ultrafast processes began to be explored by ultrafast measurements using optical, x-ray and electron pulses (8, 13–15). The birth of free electron lasers enables not only ultrafast but ultra-intense x-ray pulses for studying ultrafast dynamics. However, these early developments using relatively large beams were incapable of spatially resolving the dynamics of heterogeneities that underlie many fundamentally important and technological relevant phenomena.

Combining the techniques in time and space domains, innovative spatiotemporally resolved imaging tools (Figure 1b) have been an important technical driver to propel materials characterization into a new era. For example, ultrafast electron microscopy (8, 16–19) offers large scattering cross sections to study 2D materials (16, 20, 21). Ultrafast scanning near-field optical microscopy are available for studying nanoscale electronic processes (22–25). Ultrafast scanning tunneling microscopy can achieve ultrahigh temporal and spatial resolution probes (26).

Among many microscopic techniques, x-ray imaging techniques have been essential toolsets for nanoscale science because x-rays provide direct probes of order parameters and electronic structure in materials. Their unique capabilities include penetrating power for studying buried materials, element sensitivity for mapping chemical species, and flexible sample environment for *in-situ*, *in-operando* studies of real materials and devices. These advantages over other imaging modalities are inherited by spatiotemporally resolved x-ray imaging techniques.

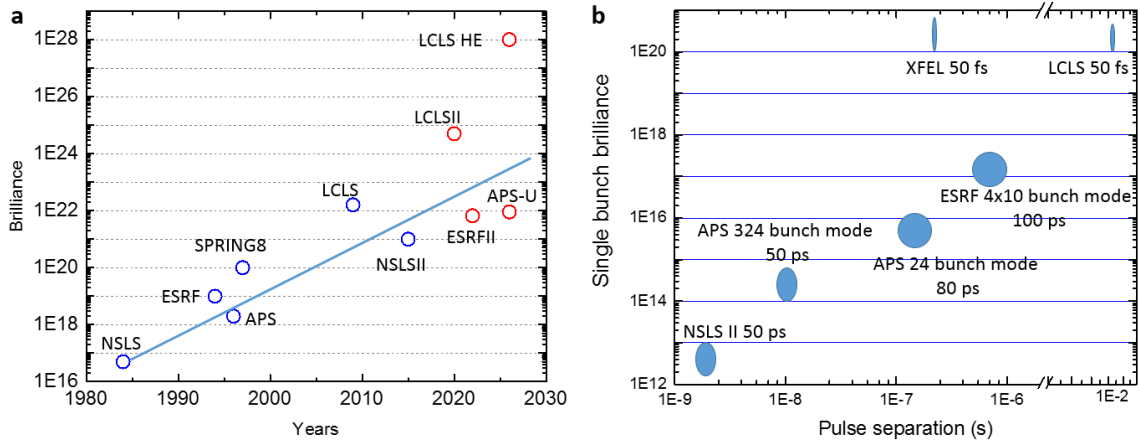

Figure 2 (a) Estimated 10 keV source brilliance at several hard x-ray facilities around the world. The red circles represent the proposed new facilities. (b) The pulse interval and single bunch brightness of various x-ray facilities. The pulse durations are quoted as full-width half maximum and indicated not-to-scale by the width of the data points. NSLS, National Synchrotron Light Source; SPRING: Super Photon Ring.

X-ray facilities around the world have been rapidly developing in recent years. X-ray brilliance at large-scale facilities, as defined by photons/s/mm$^2$/mrad$^2$/0.1%bandwidth, exponentially increases and has been accelerated by the birth of free electron lasers (FELs) (Figure 2a). Many third-generation storage ring (SR) sources are racing to upgrade to the next generation diffraction-limited source that offers x-ray radiation with smaller emittance, higher energy, and enhanced coherence (27). As a result, existing nanoscale imaging techniques will be benefted with the promise to reach a spatial resolution of 1 nm (28). Therefore, the on-going upgrade to the fourth generation synchrotron sources around the world opens unprecedented opportunities for nanoscale imaging. On the other hand, the FELs remain the best facilities to push the temporal resolution (29, 30). The combination of ultrafast capability with the improved spatial resolution has just started to be explored with novel applications. Among these x-ray sources, SRs offer high-repetition-rate operation while FELs offer high single-bunch brightness (Figure 2b). Due to the nature of the x-ray sources, they are optimized for x-ray imaging in specific time and space regimes as detailed later.

At the dawn of new large-scale x-ray facilities, this review intends to focus on time-resolved x-ray imaging, and to complement the many excellent reviews mainly on static x-ray microscopy (9–11, 31–34). We discuss recent advances in combining spatial and temporal resolution for next-generation x-ray imaging facilities and their applications in materials science. In Section 2, we first describe the principles of x-ray measurements for achieving the required spatial and temporal resolution. In Section 3, we present recent examples in applying time-resolved x-ray imaging techniques to solve various scientifically challenging problems in both fundamental and applied sciences. In Section 4, we provide an outlook of future research and discuss the opportunities as well as challenges of time-resolved x-ray microscopy.

## 2: Principles of time-resolved x-ray microscopy

To reveal the evolution of nanoscale events, simultaneous high-resolution measurements in both space and time are needed. A natural route to develop time-resolved imaging techniques is to combine well-established techniques for achieving high temporal and spatial resolution simultaneous. This new trend of research has just begun to be explored at the x-ray wavelength. In this section, we present the measurement principles used to achieve high spatial and temporal resolution. Then, we discuss the progress of combining these techniques for spatiotemporally resolved x-ray measurements.

### 2.1 Spatial resolution

### 2.1.1 Scanning probe imaging

Scanning probe x-ray microscopy uses an x-ray beam that is either focused or confined laterally to the beam propagation direction to create a set of spatially localized measurements. The contrast from these measurements then produces a real space image of the sample volume where the spatial resolution is a convolution of the beam size and scanning precision. This imaging technique flexibly enables a variety of direct imaging techniques using any of the fundamental contrast modes such as absorption, diffraction, or fluorescence. This real-space microscopy provides quantitative relationships to material properties within the scattering volume; however, the data collection is typically slower than full-field imaging approaches for a comparable field of view as each pixel in a raster map must be independently acquired. Although typical uses are overwhelmingly for 2D projected imaging methods, some of the most exciting recent applications have been the enabling of 3D tomography as vector (or multivariate voxel) tomographic renderings that are made possible by micro- and nano-focused scanning beams such as using x-ray fluorescence (35), small-angle x-ray scattering (36), or magnetic scattering (37).

X-rays can be focused using any of the major interactions of light with matter – reflection, refraction, or diffraction - creating a variety of methodologies and approaches for scanning microscopy (38, 39) with geometries that can be designed to match source parameters to a desired spot size and divergence. Grazing incidence Kirkpatrick-Baez focusing mirrors typically provide high-flux micron and sub-micron beam spots for scanning microscopy and spectroscopy (40). Fresnel zone plate diffractive optics provide focused beam spots in the tens of nanometers and are commonly used for nanoscale microscopy in both the soft x-ray (100 -1000 eV) (31) and more recently hard x-ray (7,000 – 20,000 eV) (9, 41) regimes. The limitations of spatial resolution for each focusing approach are effectively determined by how large a numerical aperture optic can be created without significant aberrations.

The most recent advances in x-ray optics for the scanning probe are the integration of multi-layer-Laue diffractive focusing optics into scanning microscopes giving access to sub 10 nm focal spot sizes for hard x-rays (42). Another advance is the use of nano-focusing optics in phase-retrieval imaging techniques to increase tolerance of voxels to phase shift and create a 3D image without sample rotation (43). Improvements of scanning speed such as fly scanning (44) and advanced visualization (32) have also greatly increased data collection rate. Multimodal diffraction microscopy such as high bandpass alignment–low bandpass imaging (45) and correlative structural / chemical imaging (46) are also more common.

**2.1.2 Full-field imaging**

Different from scanning probe imaging, full-field x-ray imaging can resolve features within the beam size. The simplest form is projection imaging, in which light propagates along a straight line to cast images that arise from the absorption or reflection contrast to the area detector. The magnification and field of view of this scheme are usually limited by the distances between the light source, the sample and the detector. For high magnification, the required large distance makes the instrument size as long as 50 meters (47). This constraint can be lifted by designing appropriate imaging systems so that large magnification can be achieved by an imaging instrument with manageable footprint, typically a few meters. In this scheme, the object is illuminated by the light source, and the reflected or transmitted light is mapped to the image plane on an area detector. A magnification of several hundreds is routinely achievable, depending on factors such as numerical apertures of the imaging system and the detector pixel size.

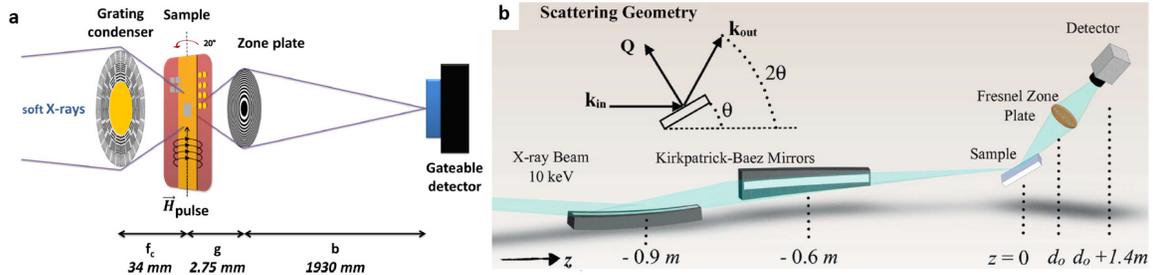

Figure 3 Full-field x-ray imaging in transmission (a) and refection (b) geometry. Reproduced with permission from *(48)* and *(49)*.

Full-field x-ray imaging can operate in the transmission or reflection mode. In the transmission geometry (Figure 3a), the contrast mechanism replies on x-ray absorption or phase contrast after passing through the sample. Since the x-ray optics do not need to be mobile in this collinear geometry, the working distance of the imaging lens can be as small as a few mm to achieve high numerical aperture for a typical resolution of 20 nm (48). In the reflection geometry (Figure 3b), the contrast mechanism is x-ray reflection or diffraction intensity, which is highly sensitive to atomic configuration. The imaging system including objective lens and detector, usually needs to be movable to follow the reflective beam. This requirement puts technical constraints on the working distance and the stability imaging optics, thus limiting the spatial resolution to the order of 70 nm (49).

**2.1.3 Coherent diffraction imaging**

Coherent Diffraction imaging (CDI) is a lensless imaging technique, which overcomes the limitation on attainable resolution imposed by x-ray optics, taking advantage of the coherence of the x-ray radiation. In CDI, the object to be imaged is illuminated by a coherent x-ray beam. The diffracted x-ray patterns usually appear as speckles, which are then measured in the far-field by an x-ray detector (Figure 4a). The diffraction intensity is proportional to the Fourier transform of the illuminated object squared, during which the phase information is lost. Both phase and intensity are required in order to calculate the inverse Fourier transform and thereby reconstruct the real space object. If, however, the diffracted intensities are sampled more finely than a factor of two (Nyquist criterion), the measured diffraction pattern uniquely encodes both the intensity and phase of the Fourier transform, and the object image can be recovered through the use of iterative phase retrieval algorithms (50, 51).

When CDI is performed in the Bragg geometry, known as Bragg CDI (BCDI), the recovered phases are related to the lattice displacement projected along the scattering vector. For instance, if the measurement is performed around 111 Bragg peak, the recovered phases are proportional to the lattice displacement projected along the [111] direction. The gradient of this displacement field is the projected lattice strain. CDI can be performed in a number of ways: conventional CDI uses a beam larger than the imaged object, typically an isolated nanoparticle or a single grain within a polycrystalline material. By contrast, x-ray ptychography (Figure 4) images multiple overlapping areas with a beam that is smaller than the imaged object (43, 52). The advantages of ptychography over conventional CDI are that the technique is not restricted to isolated samples, and the multiple overlapping constraints make the reconstructions more robust (34). The resulting spatial strain resolution is on the order of a few tens of nm, with picometer sensitivity to strain. As we will see in the following sections, this sensitivity to strain makes CDI a powerful technique to study materials response in a variety of operating conditions and under a variety of external stimuli.

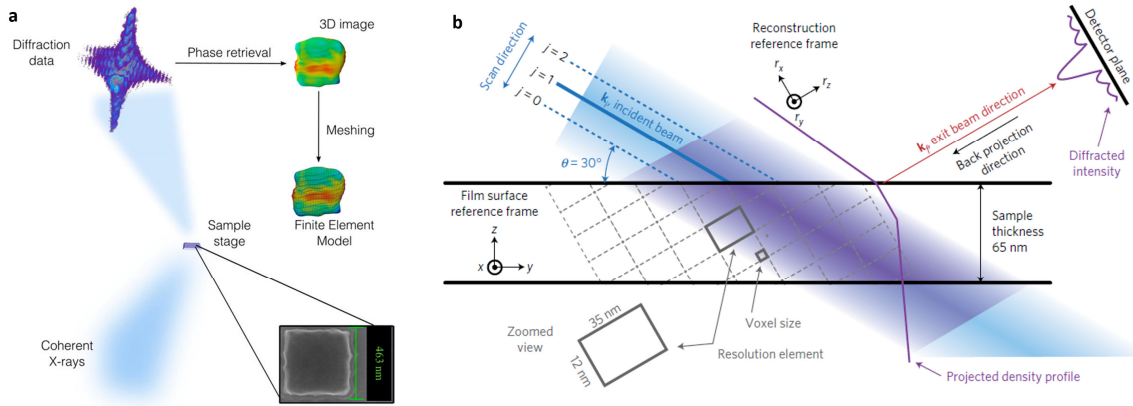

Figure 4 (a) Schematic of conventional CDI measurement in which the sample is smaller than the beam size. Reprinted with permission from *(53)*. (b) Schematic of 3D Bragg ptychography where the sample is larger than the beam size and is scanned through overlapping measurements. Reprinted with permission from *(43)*.

## 2.2 Temporal resolution

The structural dynamics in materials can span a hierarchy of time scales (Figure 5a). For example, when a material is excited by an optical pulse, a cascading series of events within the material is set in motion. The first action can be electronic excitation to promote delocalized electrons in the conduction band, provided the excitation photon energy is higher than the band gap. Energy from incident photons is imparted to the excited electrons in the form of kinetic energy, giving rise to so-called hot electrons within tens of femtoseconds. These hot electrons transfer their kinetic energy to the underlying lattice, through interactions between electrons and lattice, often referred to as electron-phonon coupling. This transfer of kinetic energy is material dependent and its time scale can vary from hundreds of fs in metals to tens of ps for certain semiconductors (54). Subsequently, collective lattice motions are initiated as optical phonon oscillates on ps time scales or as acoustic wave propagates with a characteristic time scale of ns over a μm distance. On μs to ms timescales, the heated material begins to relax via energy exchange with its surrounding enviroment. As shown in Figure 5a, these hierarchical time scales, ranging from fs to ns and from ns to ms, can be accessible by FELs and SRs.

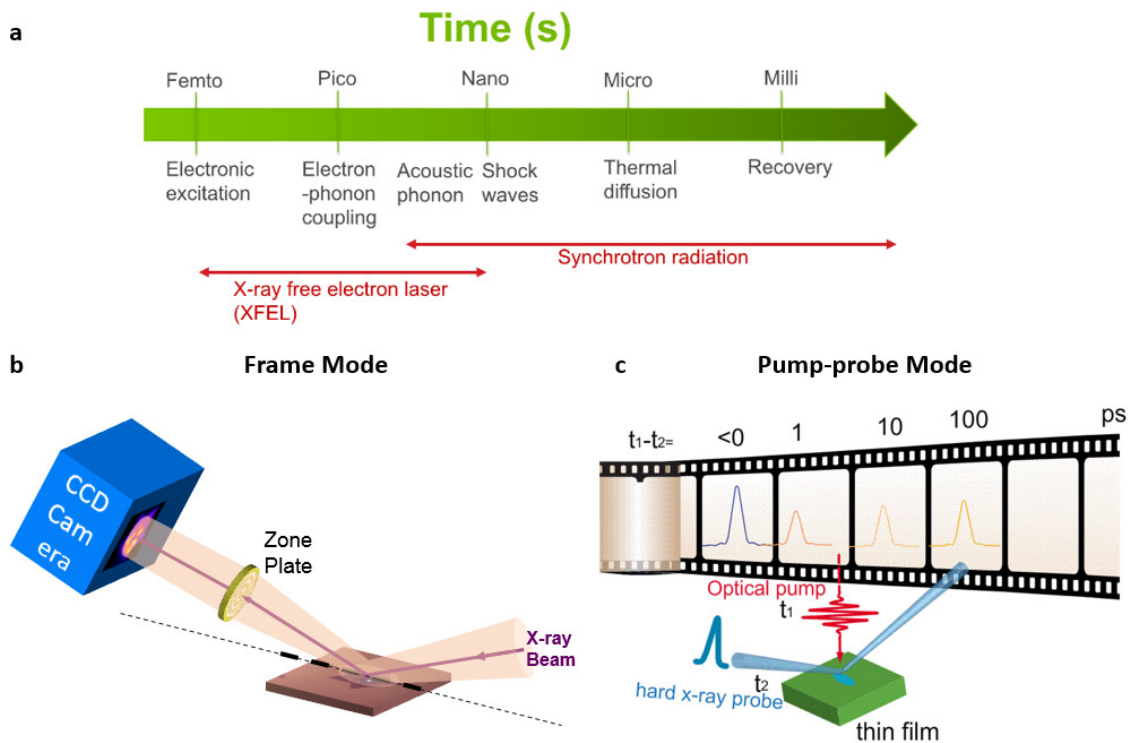

Figure 5 (a) Time scales of representative material dynamics following optical excitation, which are accessible by XFELs and SRs. (b) Frame Mode: an x-ray imaging detector is used to record time-dependent images frame by frame. (c) Pump-probe mode: a pair of pulses with controllable delay arrives at the sample to excite and probe the sample, respectively.

To enable the temporal resolution of nanoscale x-ray probes, two modes are widely used. One is "frame mode", in which the temporal resolution is related to the x-ray detector frame rate. The other is "pump-probe mode", a stroboscopic technique in which the temporal resolution is usually related to the duration of the pump and probe pulses.

## 2.2.1 Frame mode

The frame mode refers to the data collection mode to record sequential x-ray images by an x-ray camera, similar to shooting a movie (Figure 5b). In this mode, the temporal resolution is achieved by the frame rate of the detector. The fastest frame rate of an x-ray area detector is on the order of MHz, which set a temporal resolution to sub-μs (55). In this mode, the temporal resolution is much longer than the x-ray pulse interval and pulse duration, and the source such as SRs can be regarded as a continuous wave light source. Its temporal resolution is limited to image quasi-equilibrium processes from µs to minutes. For high temporal resolution faster than the movie mode, a pump-probe scheme is used.

**2.2.2 Pump-probe mode**

The pump-probe mode refers to a widely used data collection mode for ultra-high temporal resolution (Figure 5c). A pump refers to an event that stimulates the system into action. After a controllable time delay, a probe is used to characterize the resultant changes. The duration of pump-probe events is on the order of the pump and probe pulse lengths, which usually set the temporal resolution of this mode. Pump-probe events can be repeated many times at fixed time delays to gain signal-to-noise ratio, so that the temporal resolution is independent of the data acquisition speed of the detector. However, due to the stroboscopic nature of this mode, only deterministic processes that follow the same evolution pathways can be measured in repetitive measurements, while any stochastic processes are averaged to wash out by integrating the probing signal over many pump-probe events. The single-shot pump-probe has been demonstrated to overcome this limitation to study non-reversible phenomena. But it usually requires ultrahigh single pulse intensity available at FELs and replenishing samples for each of pump-probe measurements (56, 57).

**2.2.3 Pulse duration and timing structure of large-scale x-ray facilities**

The x-ray pulse duration is an important parameter for the time-resolved x-ray measurement. The x-ray pulse duration is proportional to the electron bunch length. As shown in Figure 2b, the pulse duration at SRs is usually on the order of 50-100 ps. Various schemes have been implemented to improve the temporal resolution at synchrotrons at the cost of lower x-ray flux, including reducing the bunch current thus the length of the bunch (58), reducing bunch length by slicing the electron bunch (59), and resolving the x-ray pulse temporal profile using a streak camera (60). These schemes with low efficiency are outpaced by the fast development of FELs that deliver 50 fs x-ray pulses routinely, with unprecedented single bunch flux up to $10^{12}$ photons per pulse. The temporal resolution at FELs is not usually limited to the pulse duration at FEL but the ability to achieve high precision synchronization between the pump laser and probe x-ray pulses. A timing sorting scheme that *in-situ* records the delay of each pump-probe event has been used to help achieve a few fs temporal resolution (61).

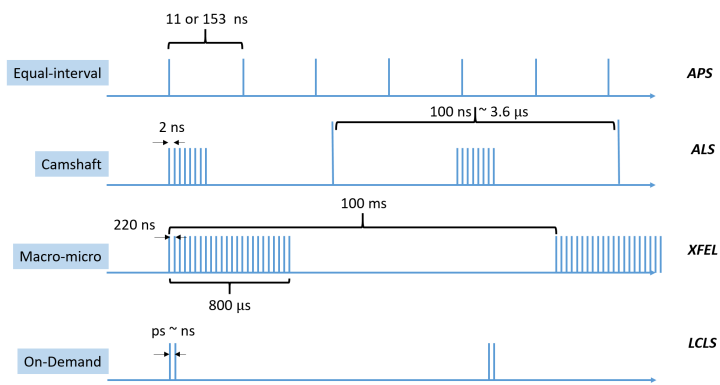

Figure 6 Various x-ray bunch structures that have been demonstrated at the representative large-scale x-ray user facilities.

> **BUNCH STRUCTURES**
> (a) Equal-interval: The x-ray bunch is equally distributed in time. The repetition rate is the inverse of the time interval of the bunches.
> (b) Camshaft: Camshaft bunch refers to a single bunch between trains of tightly spaced bunches at SRs. This bunch is usually filled with higher charge current, thus offering higher single-bunch brightness. The isolation of this bunch from others in time offers the opportunity to gate its intensity for time-resolved x-ray measurements.
> (c) Macro-micro: This bunch structure is composed of macro bunches of densely packed x-ray pulses (micro bunches). The temporal intervals of pulses within micro and macro bunches are usually on the order of ns and ms, respectively. This is usually used at linac accelerator-based light sources.
> (d) On-demand: This mode is a configurable operation mode at FELs that allows photoinjection of electron bunches into the accelerator on-demand, with flexible time intervals between bunches and controlled injection of desired bunch current.

Another important aspect of time-resolved x-ray imaging is the bunch structure at large-scale x-ray facilities. Bunch structure refers to the filling patterns of electron bunches in the accelerator, which directly determine the x-ray pulse structure. Depending on accelerators, the bunch structures of various x-ray light sources can be vastly different (Figure 6). Understanding x-ray bunch structure can help to select the appropriate pump repetition rate and the suitable detector. To achieve the temporal resolution promised by pump-probe techniques, the pump and probe need to run at the same repetition rate. Or, the probe pulse needs to be isolated during the detection, either by mechanical choppers or gatable x-ray detectors. Commonly used bunch structures are summarized in the sidebar titled "Bunch Structures".

### 3. Applications of time-resolved x-ray microscopy in materials science

One of the major goals of time-resolved x-ray microscopy is to visualize the local structural dynamics in solids. Structural dynamics describes time-dependent order parameters of lattice, charge, spin and orbital degrees of freedom. To date, the applications of time-resolved x-ray imaging are mainly focused on lattice dynamics and nanoscale magnetism. This is because the existing light source can provide enough flux for imaging lattice and magnetic heterogenuities, while time-resolved x-ray imaging of other local order paramters may become feasible with further the improvement of the light source and techniques.

Lattice dynamics describes time-dependent atomic displacement from static positions. Depending on the consequence of atomic displacement, structural dynamics can be categorized into linear and nonlinear regimes. In the linear response regime, the structure symmetry is largely maintained. Structural change only accommodates the energy transport such as elastic strain wave propagation but does not consume energy to change the symmetry of the structure. In the nonlinear response regime, atomic displacement is large enough to lead to structural symmetry changes, such as monoclinic to tetragonal phase in $VO_2$. The materials structural properties can be significantly changed which in turn drives other functional properties such as metal-to-insulator phase transition.

Spin is a fascinating degree of freedom that gives rise to a wide range of dynamical phenomena in nanoscale magnetism, including ferromagnetic domains, magnetic vortices, and topological structures such as magnetic skyrmions. Motivated by understanding fundamental processes and applications of data storage and processing using spins, time-resolved x-ray microscopy has been an essential tool to study dynamics of nanoscale magnetism because it offers higher spatial resolution and element sensitivity, comparing with time-resolved magneto-optical Kerr effect microscopy (62). Direct observation of nanoscale magnetic dynamics is one of the most successful applications of time-resolved x-ray imaging.

**3.1 Evolution of local strain**

**3.1.1 Imaging dynamical strain field in nanocrystals**

As described in Section 2, Bragg CDI (BCDI) provides both structural information with ~10 nm resolution and crystallographic information with pm sensitivity, painting a unique picture of crystallographic deformation at the nanoscale. Following an external excitation of the material, such as mechanical loading, temperature change, catalytic activity, etc., local structural dynamics as fast as ps can be measured using the pump-probe technique. At each temporal offset between the pump and the probe, a full 3D rocking curve is measured and as a result, 3D snapshots are obtained for the time-varying strain fields within materials following optical excitation.

The first successful demonstration of time-resolved BCDI measurements was done by Clark et. al. (64), who measured the time-varying displacement field in isolated gold nanocrystals with a size of 300 to 400 nm after optical excitation. Figure 7a shows the change of diffraction fringes comparing the coherent diffraction patterns of 111 Bragg peak 10 ps before and 60 ps after the excitation. Figure 7b shows the angular shift of the Bragg peak shows oscillatory behavior. Clark et al. fit the observed response to two oscillatory modes, assuming the two-temperature model that energy transfers between hot electrons and cold lattice. The measured structural responses of two crystals show vastly different dynamic due to their different sizes and shapes.

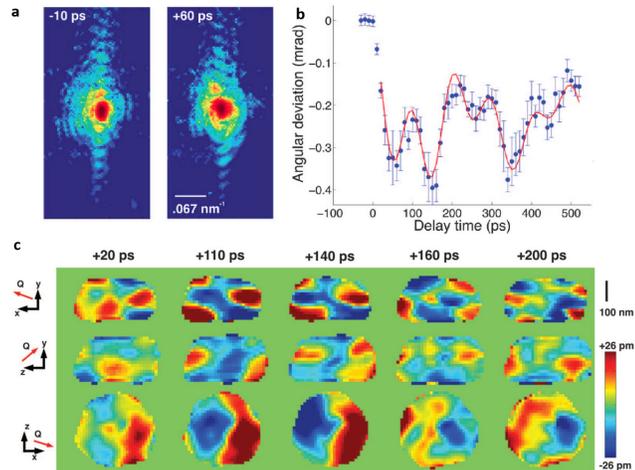

Figure 7 (a) Coherent diffraction patterns recorded at delay of -10 and 60 ps. (b) The angular deviation of the (111) Bragg peak as a function of time. (c) shows 2D slices through the reconstructed object and displacement field. Reproduced with permission from *(63)*.

This highlights the importance of interrogating the response of individual particles as opposed to measuring an ensemble response.

In addition to deformation modes, as inferred from the angular deviation of the Bragg peak, multiple shear modes are observed with much higher frequencies. The evidence for these modes was seen in the recovered lattice displacements within the crystal at a different temporal offset from the incident laser. Figure 7c shows the lattice displacement along orthogonal slices through the crystal at a few snapshots in time. Interestingly, while the extent of deformation of the crystal from the long-time oscillations was around 600 pm, the amplitude of the higher frequency oscillations was ~ 50 pm. It would be extremely challenging to characterize such weaker, higher frequency crystal deformation modes purely from the angular deviation of the Bragg peak, and this underscores the importance of the sensitivity to local strain provided by BCDI.

More recently, Cherukara et al. (65, 66) have demonstrated 4D time-resolved BCDI measurements at a syncrotron, the Advanced Photon Source. By choosing suitable material systems, the temporal resolution at synchrotrons is sufficient to resolve the structural evolution. In this study, the authors measured the time-dependent deformation of an isolated crystal of ZnO on a SiO$_2$ substrate. ZnO has been used to make nanowire arrays that can convert mechanical energy into electrical energy, potentially at the nanoscale (67). Consequently, being able to visualize the deformation of individual ZnO nanopillars is vital to the improved design of nanomechanical energy harvesters based on these materials.

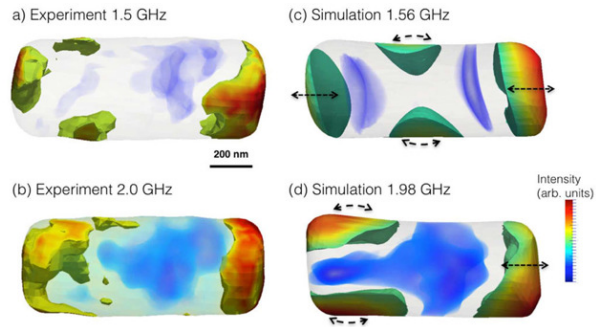

Figure 8 Temporal Fourier transform of the time-varying displacement field within the crystal. Images show the relative contribution of different portions of the crystal to oscillations of a given frequency (1.5 GHz and 2.0 GHz). Regions in red oscillate strongly at that frequency, while regions in blue oscillate weakly at that frequency. Reprinted with permission from (65).

After optical excitation, the dynamics of the crystals were measured by time-resolved BCDI of 002 Bragg peak. Since the lattice displacement in BCDI measurements is only recorded along one crystallographic direction, the measured data set is a mix from the contributions of different axial, radial and torsional modes. To better understand different modal contributions to the measured displacement field, Cherukara et al. took a temporal Fourier transform of the time-varying displacement field within the crystal. The relative contributions of different portions of the crystal to the deformation mode at 1.5 GHz and 2.0 GHz are shown in Figure 8, in agreement with the simulated results. The ability to image such inhomogeneous high-frequency deformation modes within the nanocrystal has implications for the design of improved nanogenerators, and the authors used the experimentally informed model to predict that nanorods employed in torsion would generate a 50% higher electric potential than if the bending mode were used, as is typically the case (65).

### 3.1.2 Domain dynamics in ferroelectrics

Moving beyond a single nanoscale object, many functional materials possess intrinsic heterogeneities known as domains. Ferroelectrics is one of the prominent examples of functional materials with domains (68). Ferroelectricity originates from structural symmetric breaking that gives rise to a spontaneous electric polarization. However, the alignment of polarization is usually confined into nm to μm scales, forming ferroelectric domains. Different domains have different polarization states, as well as different associated local dynamics. These complications limit the fundamental understanding and applications of pivotal ferroelectric processes such as polarization propagation, switching, and coalition.

The early studies of the structural dynamics of ferroelectric domains have focused on the structural response driven by the pulsed electric field. In the pioneering work, Grigoriev et al. (69) applied the electric field with a pulse width of several hundreds of ns, and monitored the Bragg reflections of the tetragonal phase of Pb(Zr)TiO$_3$ film by focused x-ray beam. They found the polarization switching time to be inhomogeneous in the homogenous driven electric fields, suggesting localized intrinsic sample response. By tracking the switching time as a function of position, they measured the domain wall velocity to be 40 m s$^{-1}$.

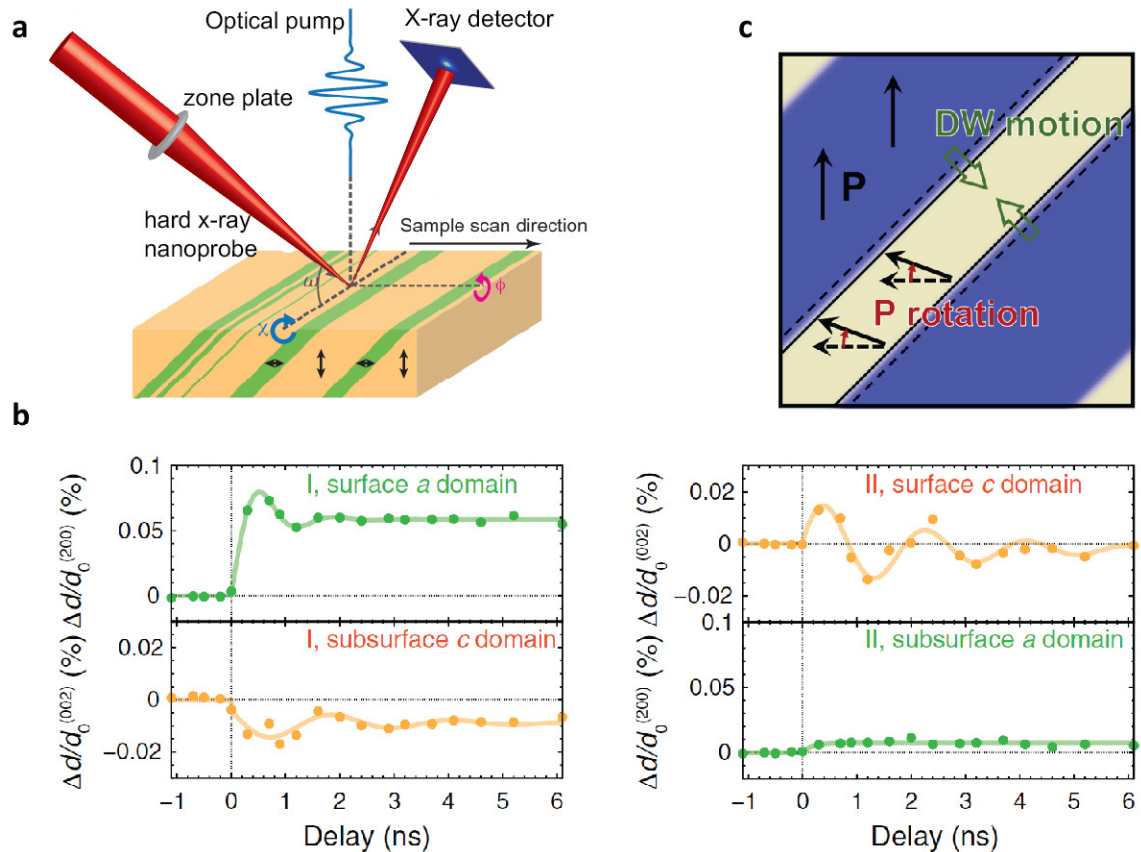

Figure 9 (a) Laser-pump, x-ray micro-diffraction probe of ferroelectric domains in BaTiO$_3$ single crystal. (b) Strain as a function of time in various representative domains. (c) Derived domain wall and polarization dynamics based on dynamical phase field simulation. Reproduced with permission from *(70)*.

Faster ferroelectric domain dynamics need to be initiated by an impulsive excitation with a duration shorter than the characterized time scales of interest. Delivering such short electric fields using electrodes on ps time scales is a challenge. Alternatively, photoexcitation is an effective way to couple energies to materials on an ultrafast time scale via electronic excitation. Ferroelectrics are excellent light responders as manifested by strong photostriction (71–73) and photovoltaic effects (74, 75). However, how light interacts with individual ferroelectric domains on nanoscales is not yet known until the following studies.

Using laser pump and x-ray microdiffraction probe at 7ID-C beamline of the APS, Akamatsu et al. (70) measured the response of individual domain in bulk $BaTiO_3$. $BaTiO_3$ is a prototypical ferroelectric material with intrinsic ferroelectric domain size on the order of μm, with the polarization aligned along either the in-plane (*a*-domain) or out-of-plane (*c*-domain) direction (Figure 9a). Upon optical excitation, they independently measured surface domains and sub-surface domains, which can be either *a* or *c* domains, by recording time-dependent 200 and 002 Bragg peaks, respectively. First, the oscillation of Bragg peak positions was observed as a result of periodical lattice contraction and expansion. The oscillation periods agreed with the round-trip time of gigahertz acoustic wave propagating between domain walls. Second, the magnitude of the domain response depends on the domain location with respect to the surface. The response of the surface domains is much stronger than that of the sub-surface domains as shown in Figure 9b. The larger surface domain response was induced by the emergence of a large surface electric field on the order of 1 MV/cm, as supported by the dynamical phase-field simulation. From the dynamical response of the neighboring domains, the speed of domain wall motions is derived to be 2.5 m / s (Figure 9c). This direct probe of structural response upon optical excitation reveals domain-dependent acoustic modes in ferroelectrics which allowed a direct comparison with phase-field modeling.

In addition to intrinsic heterogeneities, time-resolved x-ray diffraction also can help understand the energy transport in extrinsic hoterogenuities, the fabricated structures in nanoscale devices. In the recent work, Zhu et al. (76) used metasurface structures on the $BaTiO_3$ thin film to enhance the incident terahertz (THz) fields locally. Real-space time-resolved x-ray diffraction microscopy following THz excitation reveals the local and transient strain distribution around a meta-surface device. The measured strain profile at 100 ps showed a significant broadening, in in good agreement with the prediction of a ballistic-phonon-transport model but in poor agreement with a simulated profile using a diffusive model, suggesting a ballistic phonon transport process. The measured ballistic phonon transport was over a distance of hundreds of nm, two orders of magnitude longer than the averaged phonon mean free path in $BaTiO_3$, due to resonant excitation of low-frequency phonon modes. The demonstrated time-resolved real-space visualization of phonon dynamics opens up opportunities to engineer and image nanoscale transient structural states with new functionalities.

## 3.2 Dynamical phase separation

Dynamical phase separation during a solid-solid phase transition poses a great challenge for understanding the fundamental processes in phase-change materials. The early application of time-resolved imaging with a relatively poor spatial-temporal resolution of 2.5 μm and 0.7 s is sufficient to visualize dendritic growth in binary alloys (77), a classical phase transformation phenomena. However, to interrogate more volatile phase transformation, higher temporal and spatial resolutions are needed. Muli-frame ultrafast x-ray imaging with time resoltuion on μs to ms has been developed to study shock compression (78), high-rate loading (79), reactive sintering (80), additive manufacturing (81). While x-ray pump-probe imaging has been applied to study microscopic phase transformations on ps to ns time scales (82, 83).

**3.2.1 Real-time monitoring of additive manufacturing**

Multi-frame ultrafast x-ray projection imaging has been applied to *in-situ* visualize laser powder bed fusion, an important additive manufacturing process. Additive manufacturing involves melting and solidification, usually upon excitation by high-power lasers. Although it is a conceptually simple practice, the optimization of the process requires a fundamental understanding of its highly dynamic and heterogeneous nature. High-energy x-ray imaging based on full-field fast-frame imaging provides the required penetration power and spatial and temporal resolutions for studying additive manufacturing.

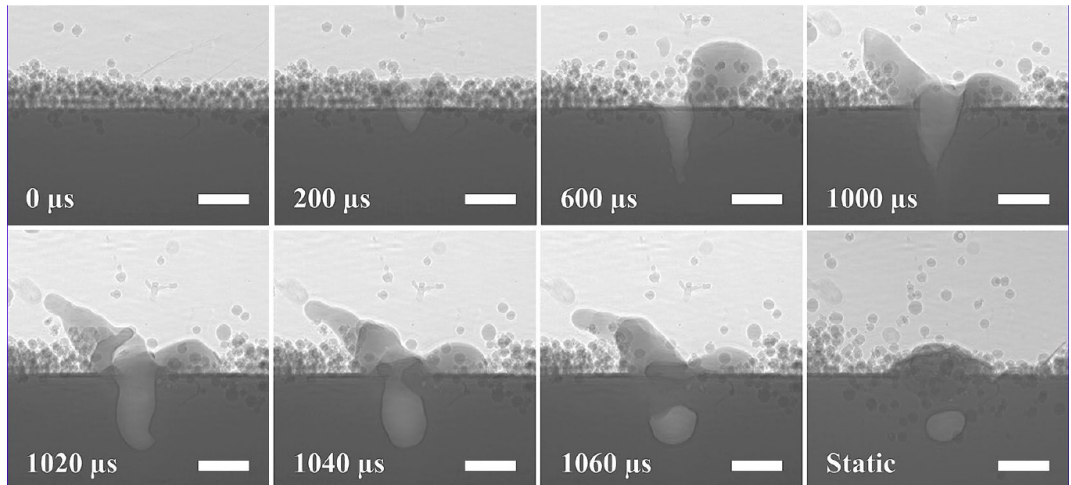

Figure 10 Multiframe x-ray images of laser powder bed fusion processes in the spot welding mode as a function of time. The high power laser strikes from the top to the middle of the powder surfaces at time zero with a spot size 220 μm. The scale bar is 200 μm. Reproduced with permission from *(81)*.

Zhao et al. (81) demonstrated the *in-situ* imaging of metal 3D printing at the APS (Figure 10). The 1070 nm continuous-wave laser was focused through a viewport into the samples in a vacuum chamber. The transmitted x-ray beam at 24.4 keV through the sample area was converted to visible light by a scintillator for imaging. Before the laser excitation, the powder bed and the base plate were clearly revealed. During the illumination of the laser, a violent melting of the powder and base was initiated. Right under the laser beam, the sample was partially vaporized and a cavity was developed with oscillatory behaviors governed by the competition between the Marangoni convection and the recoil pressure. After the laser was turned off at 1000 μs, the melted substances started to cool down and solidify into a dome shape structure. Due to the sudden collapse of the melt near the sample surface, a large pore was trapped inside the sample, also known as a keyhole pore. These spatiotemporally resolved measurements give many microscopic insights including melt pool dynamics, powder ejection, rapid solidifications, and phase transformation during additive manufacturing.

### 3.2.2 Melting and solidification of nanocrystals

As shown in the previous examples of additive manufacturing, x-ray frame imaging is able to study relative slow processes (μs to ms) while the intrinsic melting process occurs on faster time scales cannot be captured. Using ensemble-averaged ultrafast x-ray diffraction, nonthermal melting on ps time scales have been observed (84, 85). But the microscopic processes, which can depend on the size and shape of local crystal structures, are not yet known. Time-resolved BCDI measurements are able to reveal microscopic and size-dependent melting and solidification. In a recent study, Clark et al. (82) used BCDI to image the 3D structure of gold nanoparticles when heated by a strong optical pulse that leads to the partial melting of nanoparticles. They observed a transient melting of the surface of the nanoparticle, while the bulk of the nanoparticle remained solid, an observation that is consistent with the core-shell model of nanoparticle melting. Figure 11 shows the temporal evolution of a gold nanoparticle of 300 nm in diameter as a function of time for three laser fluences. Significant surface melting manifested as the low density "halo" surrounding a smaller size core is seen within ~50 ps. This result confirms a molecular dynamics simulation that shows partial melting at the nanoscale in a nonhomogeneous fashion, providing insight into the phase transition in nanoparticles.

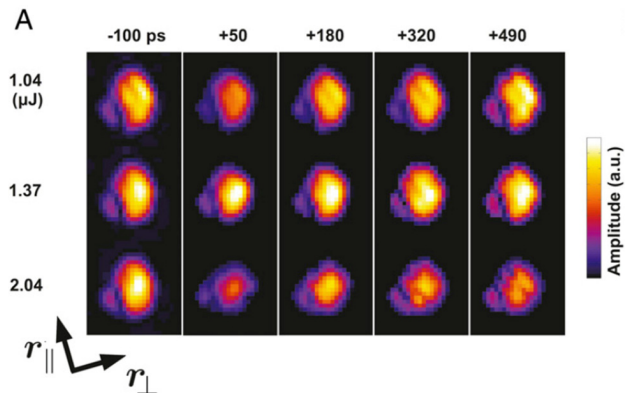

Figure 11 Transient surface melting of gold nanocrystals imaged through pump-probe BCDI measurements. Snapshots show the reconstructed crystal shape over time for 3 different laser pump pulse energies. Reproduced with permission from Ref. *(82)*.

### 3.2.3 Structural phase progression in photoexcited $VO_2$

Structural phase transition can be intimately tied to properties of materials. For example, $VO_2$ exhibits a distinct structural phase transition from the monoclinic (M) to rutile (R) lattice structure, in accompany with orders of magnitude change of conductivity known as a metal-to-insulator phase transition (MIT). This transition can be triggered by optical excitation on ultrafast time scales, holding promises for low-power ultrafast optoelectronics. Although ultrafast microscopic studies of electronic phase transition indicate the heterogeneous photoinduced MIT (22), little is known about the associated microscopic structural phase transition.

Zhu et al. (83) studied the photoinduced structural phase transformation by imaging the evolution of distinct structural phases in real time, providing critical information between the onset ultrafast phase transition that occurs on ps time scales (86–88) and the completion of the phase transformation that occurs on the ns time scale (89). In this experiment, the diffraction intensities of M and R phases were monitored. Time-dependent spatial maps were collected by raster scanning the sample against a focused x-ray probe beam with a spot size of 350 nm at various delays between the pump and probe pulses (Figure 12a). After optical excitation, in addition to the ensemble-averaged diffraction intensity exchange from M to R phases, Zhu et al. found that the emergence of R phase occurs at discrete locations, e.g., site A in Figure 12b, followed by the progression of the R phase into the M phase region. The propagation speed of the R-M phase front can be quantitatively characterized. A line-cut of the R phase diffraction intensity through site A, shown as the red dashed line in Figure 12b, is plotted as a function of delay in Figure 12c.

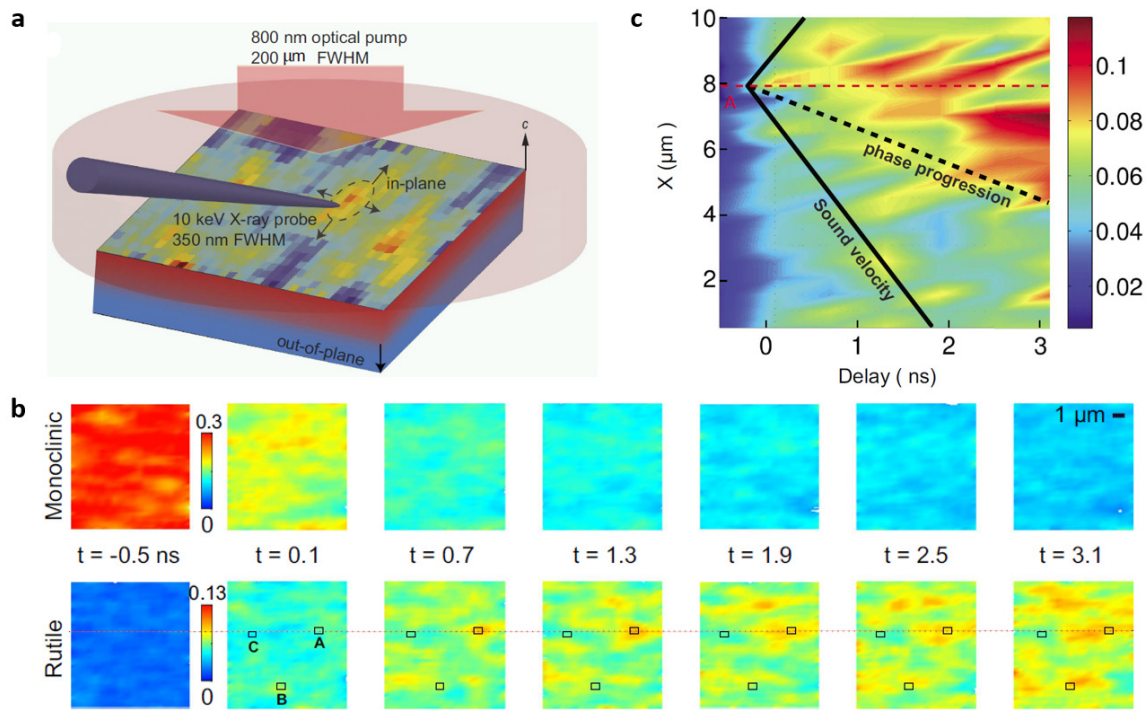

Figure 12 (a) The structural phase progression of a $VO_2$ film as probed by synchrotron-based focused x-ray pulses upon optical excitation. The blue and red regions represent monoclinic (M) and rutile (R) phases, respectively. (b) The intensity maps of the $40\bar{2}$ M phase and 002 R phase measured at a sequence of delays upon photo-excitation. (c) Space-time map of the diffracted intensities from the R phase, which is formed by the intensity line-cut along the red dotted line in (b) at various time delays. The black dashed line shows the averaged phase boundary. The solid black lines indicate the speed of sound in $VO_2$ and the red dashed line is a reference for non-propagating features. Reproduced with permission from Ref. *(83)*.

The R phase expands laterally following its initial appearance at site A with a velocity of $1100 \pm 300$ m/s, which was extracted from the slope of the phase boundary (dashed black line in Figure 12c). This speed is lower than the sound velocity of ~4 km/s but higher than the thermal diffusion speed in VO$_2$, highlighting a mesoscopic nature of structural phase transformation. Future improvements of spatial and temporal resolution will help reveal the structural details of this mesoscale phase progression process, for example, to provide direct evidence on the roles of grain boundaries and defects during MIT.

### 3.3 Dynamics of nanoscale magnetism

Nanoscale magnetism is an active research area to study the magentic structures that are confined on nanoscales at which the governing physics can differ from bulk counterpart. Although optical probes such as time-resolved magneto-optical Kerr effect microscopy can achieve ps time-resolution (62), its spatial resolution is limited to the diffraction limit of the light pulses on the order of a micrometer. At x-ray regime, the spatial resolution of the measurement can be further enhanced to a few nm due to its short wavelength. X-ray magnetic dichroism (90), an x-ray absorption spectroscopic technique, offers a direct probe of magnetic anisotropies and provides the contrast machenism for nanoscale imaging of spin configurations. Since the x-ray absorption L-edges of magnetic elements, such as Fe, Co, and Ni, are in the soft x-ray range, x-ray imaging of magnetism usually employs soft x-ray radiation.

Similar to hard x-ray imaging, various imaging techniques have been employed in the soft x-ray regime, including scanning transmission x-ray microscopy (91), full-field imaging (48), x-ray holography (92–94), and x-ray photon emission electron microscopy (X-PEEM) (95). Most of these techniques are employed in the transmission geometry. A typical setup of full-field x-ray transmission imaging at PETRA III is shown in Figure 3a. An exception is X-PEEM, which works in a "reflection" geometry in the sense that the photoelectrons released from the sample surface by x-ray radiation are imaged onto an area detector (95).

The x-ray probe can be coupled with an external excitation source to perform pump-probe x-ray imaging. The excitation of the magnetic system were achieved by delivery of a current pulse to the sample, either used for spin-pumping (96, 97) or generating a magnetic field at the sample (48, 98). The rise time of the current pulse is on the order of several hundreds of ps, which matches the temporal resolution at the synchrotron-based x-ray source. Shorter excitations using laser pulses are also avaiable (99).

Time-resolved x-ray imaging of magnetic systems has been mainly focused on two types of nanoscale structures: magnetic vortices and skyrmions. Since these magnetic structures are on the order of 10 nm to μm, a nanoscale imaging technique is required to visualize the dynamical processes of driven magnetic vortices or skyrmions.

### 3.3.1 Dynamics of magnetic vortices

Magnetic vortices in confined two-dimensional structures such as micrometer-sized disks and squares arise from self-arrangement of in-plane magnetization parallel to the edges so that the center magnetization is forced to be out-of-plane. The central region with defined out-of-plane magnetic polarization on the order of a few nm is correlated with the chirality of the vortex, as defined by the curling directions of the magnetization around the edges.

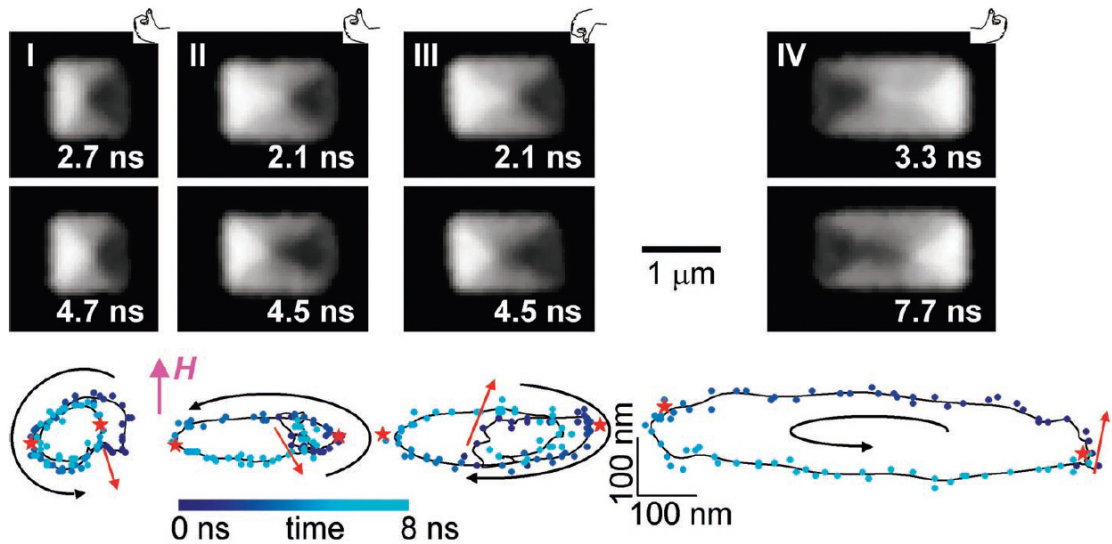

Figure 13 (Top) Domain images of the in-plane magnetization of Pattern I, II, III, IV taken at the specified delay times after the magnetic field pulse. Hands illustrate the vortex handedness and the out-of-plane core magnetization as determined from the vortex dynamics. (Bottom) Trajectories of the vortex core. The dots represent sequential vortex positions. Lines represent time-averaged positions. The progression in time is symbolized by the dot color. Red arrows show the trajectory during the field pulse; blackarrows show the direction of gyrotropic rotation after the pulse; and red stars show the vortex position for the shown domain images. Reproduced with permission from Ref. (100).

Time-resolved x-ray imaging has been a major tool to visualize the dynamics of magnetic vortices. A prominent example is the discovery of magnetic vortex gyration. In this pioneering work, Choe et al. (100) imaged the magnetization dynamics of ferromagnetic vortices in permalloy squares by time-resolved X-PEEM. The magnetic vortex exhibits a gyrotropic motion upon a fast (300 ps, full-width-half-maxium) in-plane magnetic field pulse excitation (Figure 13). The careful study of the vortex motion showed that the direction of gyration is determined by the chirality of vortices with the right-hand rule, rather than the applied magnetic field direction. The gyration period is measured to be on the order of ns. Combining the real-space tracking of the core position, the core speed can be quantitatively measured, on the order of 100 m/s. This work sparked interests in using time-resolved x-ray imaging to study magnetic vortices. The subsequent studies revealed rich details of the vortex dynamics including the eigenmodes of the gyration (101), non-gyro motion (102), free vortex motion (103), nonlinear effects in vortex dynamics (104), coupled oscillations (105), and switching of the core polarization (106–108).

### 3.3.2 Dynamics of magnetic skyrmions

Skyrmion is another magnetic structure under intense studies because of its nontrivial topological properties (109). This particle-like object has a size ranging from 1 nm to 1 μm, depending on the generation mechanims. For example, if skyrmion formation is due to the interplay of Dzyaloshinskii-Moriya interaction with the symmetric exchange interaction, a helical spin structure with well-defined handedness is formed with a size of tens of nm. One of the attractive features of skyrmions is the topological protection that prevents them to be disassembled so that they are more tolerant to geometric irregularities. In addition, the prominent motion of skyrmion can be initiated under the ultralow current density, which promises low-power consumption for information processing. Using pump-probe x-ray holography, Büttner et al. (94) reveals the gyrotropic trajectory of a skyrmionic magnetic bubble upon the magnetic field excitation through the application of current pulses (Figure 14a&b). Two gigahertz gyrotropic modes associated clock-wise and counter-clock-wise motions are identified, as needed to fit the measured trajectory that shows gyration motion (Figure 14c). The quantitative measurements of these frequencies allow the experimental determination of the inertial mass of skyrmions, which is much larger than the theoretical predictions. More recent x-ray imaging work shows that the dynamics of magnetic skyrmions can be controlled by the spin-orbit torque on the nanosecond time scale (110), and skyrmion Hall effect was confirmed by tracking the real-space trajectories of current-driven skyrmions (111).

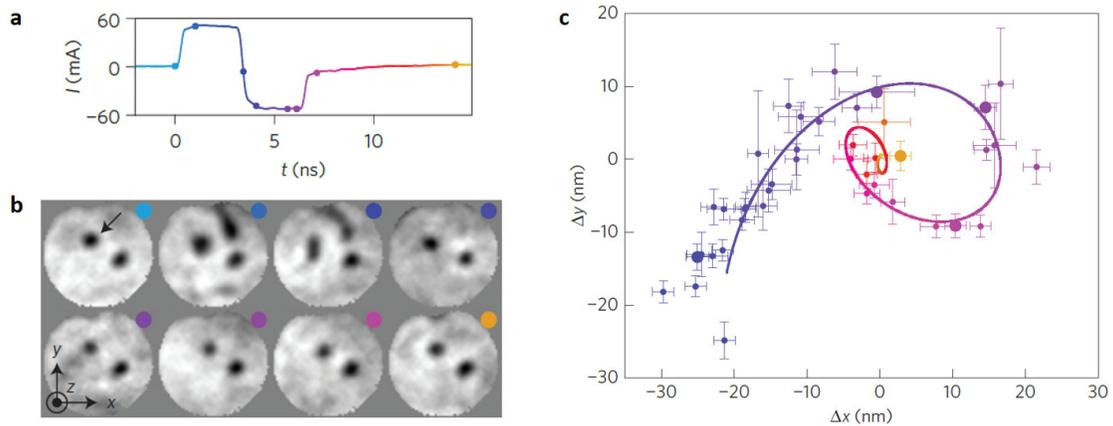

Figure 14 (a) The injection current pulse that is used to generate magnetic field pulse. (b) Magnetic domain configurations of the disk at various time delays indicated by the corresponding color dots in (a) and (c). The trajectory of the center of magnetization of the magnetic bubble, skyrmion. Reproduced with permission from (94).

**4: Outlook**

Time-resolved x-ray microscopy is gaining momentum in the race for versatile high-resolution measurements to decipher the complex dynamics of nanoscale phenomena. This will become increasingly relevant for a broad class of scientific cases where an understanding of *local* and *dynamical* response in real space is key to harnessing high-value collective phenomena. This theme is generally common to systems whose properties are driven by broken symmetry – either intrinsically such as subtle local perturbations nucleating a phase transition and local lattice strain defining the quantum behavior of a single atom defect, or extrinsically such as nanoscale engineering used to enhance confined plasmonic response and hybridize opto-acoustic resonant systems. These real-world systems are in many cases difficult to model theoretically due to miniature size or heterogeneity. Understanding complex nanoscale phenomena (see sidebar titled

"Examples of nanoscale phenomena") will be benefited greatly by direct experimental visualization of the dynamical response of heterogeneous systems.

> Examples of nanoscale phenomena
> - Nanoscale energy conversion and transport
> - Dynamics of domains and defects
> - Dynamical phase separation during phase transformation
> - Local structural evolution in quantum materials
> - Heterogeneities far away from equilibrium

To fulfill the potential of time-resolved x-ray microscopy, the continuously improvement of spatial and temporal resolutions will in large part be driven by the improvement of x-ray source properties and end-station instrumentation in both storage rings (SRs) and free electron lasers (FELs). Future SRs will provide x-ray pulses with several orders of magnitude higher brightness and coherence to greatly improve nanoscale imaging resolution and data rate, with temporal resolution suitable for studies of structural dynamics longer than the x-ray pulse duration of ~100 ps. The higher brightness will enable interrogation of heterogeneous orders with weaker x-ray scattering cross-sections, such as charge density waves and antiferromagnetic orders. The higher coherence will enable new hybrid techniques such as time-resolved x-ray ptychography for achieving high spatial resolution by exploiting coherence of the x-ray beam. At FELs, the planned upgrade of FELs (113) will provide high repetition rate and ultrashort x-ray pulses with enhanced position and spectral stability suitable for microscopic studies.

X-ray microscopy in the time domain with improved brightness, coherent flux, and repetition rate, unavoidably yields multi-dimensional data with unprecedented high rate. The parameter spaces associated with x-ray microscopic becomes very extensive, including 3D real space, 3D reciprocal space, multi-dimensional spectra, and time. The data storage, memory and computing resources required to analyze massive data sets are expected to severely tax the compute infrastructure at future light sources (114). How to analyze these data efficiently is a mounting task for time-resolved x-ray imaging. Several approaches have been taken to tackle this challenge for static x-ray microscopy (115). Non-traditional analysis techniques that leverage recent developments in deep learning and data mining (116–118) are expected to complement traditional image reconstruction algorithms especially when rapid or real-time experimental feedback is required.

The potential of time-resolved x-ray imaging for future material studies should be carefully viewed in the context of dose limitations due to radiation damage which can be in some systems comparable to electron beam damage when calculated on the basis of imaging contrast per dose (119, 120). X-ray beam damage can be caused by a variety of interactions with an ionizing beam – free radical generation, oxidation, chemical activation, differential charging, or local heating – leading to a loss of mass or loss of crystallinity in a material (121). It has been attempted to overcome these limits by using coherent diffraction imaging techniques, which remove the need for a nano-focused beam and make an explicitly phase-sensitive measurement by using the scattered intensity distribution over a wide range of momentum transfer to iteratively reconstruct a real-space image (122). However, resolutions even with coherent imaging techniques have not yet been demonstrated past the theoretical limit of direct imaging techniques discussed above in the example of biological materials (123, 124), presumably as the calculated damage limits based on fundamental contrast efficiency which remains the same regardless of imaging methodology. Some of these pathways have been shown to be unexpectedly present even in more dose-tolerant semiconductor materials using hard x-rays (125).

In parallel with the development of time-resolved x-ray imaging, ultrafast electron microscopy (8, 16–19) has also garnered recent interest. Ultrafast electron microscopy is suitable for studying in-plane suspended thin samples such as 2D materials in a forward diffraction geometry, while time-resolved x-ray microscopy is optimized for probing buried materials on substrates in a reflection geometry with sensitivity to the out-of-plane lattice dynamics. Their unique capabilities are complimentary to each other with shared scientific interests. Collaborative efforts from both community of ultrafast electron and time-resolved x-ray microscopy will be beneficial to advance the studies of nanoscale structural dynamics in materials.

In summary, time-resolved x-ray microscopy has been demonstrated as an essential tool to characterize nanoscale dynamics in materials. The increasing number of planned instruments with compatible time and space resolution around the world will meet the research needs for a wide range of research disciplines. Future development at upgraded large-scale x-ray facilities will advance the spatiotemporal resolution to reveal nanoscale processes in real time and provide predictive control of material properties with high time and space precision.

**Acknowledgment**

We thank Dr. Ross Harder, Dr. Tao Sun and Dr. David J. Keavney for the insightful discussions. H. W. acknowledge the support of U.S. Department of Energy (DOE), Office of Science, Basic Energy Sciences (BES), Materials Sciences and Engineering Division. M. C. and M. H. acknowledge the support of the Center for Nanoscale Materials, a U.S. Department of Energy Office of Science User Facility, and supported by the U.S. Department of Energy, Office of Science, under Contract No. DE-AC02-06CH11357.